\documentclass[preprint,12pt]{elsarticle}

\usepackage{amssymb}

\usepackage{amsmath, mathrsfs, dsfont}
\usepackage{lmodern,bm}
\usepackage{amssymb}
\usepackage{amstext}
\usepackage{braket}
\usepackage{algpseudocode}
\usepackage{verbatim}
\usepackage{todonotes}
\usepackage{footnote}
\usepackage{multirow}
\usepackage{ulem}
\usepackage{hyperref}

\definecolor{darkblue}{rgb}{0,0,.6}
\definecolor{darkred}{rgb}{.6,0,0}
\definecolor{darkgreen}{rgb}{0,.6,0}
\definecolor{red}{rgb}{.98,0,0}

\def\ssmall{\fontsize{8pt}{2pt}\selectfont}
\usepackage{listings}
\usepackage{caption}
\DeclareCaptionFormat{listing}{\rule{\dimexpr\textwidth+17pt\relax}{0.4pt}\par\vskip1pt#1:#2#3}
\lstnewenvironment{cpplisting}[1][]
  {\lstset{language=C++,numbers=right,#1}}{}

\lstnewenvironment{bashlisting}[1][]
  {\lstset{language=bash,numbers=right,#1}}{}

\lstnewenvironment{outputlisting}[1][]
  {\lstset{language=html,numbers=right,#1}}{}

\newcounter{bla}

\journal{Computer Physics Communications}

\begin{document}

\begin{frontmatter}

\title{Exact Diagonalization library for quantum electron models.}


\author[a,b]{Sergei Iskakov\corref{author}}
\author[b,c]{Michael Danilov}

\cortext[author] {Corresponding author.\\\textit{E-mail address:} iskakoff@q-solvers.ru}
\address[a]{Department of Physics, University of Michigan, Ann Arbor, Michigan 48109, USA}
\address[b]{Theoretical Physics and Applied Mathematics Department, Ural Federal University, Mira Str.19, 620002, Yekaterinburg, Russia}
\address[c]{Institute of Theoretical Physics, University of Hamburg,Jungiusstra{\ss}e 9, 20355, Hamburg, Germany}

\begin{abstract}
We present an exact diagonalization C++ template library (EDLib) for solving quantum electron models, including single-band finite Hubbard cluster and multi-orbital impurity Anderson model. The observables that can be computed using EDLib are single particle Green's functions and spin-spin correlation functions.
This code provides three different types of Hamiltonian matrix storage that can be chosen based on the model.
\end{abstract}


\end{frontmatter}



{\bf PROGRAM SUMMARY}

\begin{small}
\noindent
{\em Program Title: } EDLib  \\
{\em Licensing provisions:} MIT  \\
{\em Programming language:} C++, MPI  \\
{\em External routines:} ARPACK-NG, ALPSCore library\cite{Gaenko2016}  \\
{\em Nature of problem:}\\
Finite Hubbard and Anderson models play an essential role in the description of strongly correlated many-particle systems. These models consist of a small number of localized orbitals with Coulomb interaction between electrons and (in case of the Anderson model) non-interacting bath energy levels.
The finite Hubbard cluster can be used to study molecular magnets, such as $Mn_{12}$, $Fe_4$, $Mn_4$, and $V_{15}$, which are currently of interest due to their potential for use in novel technologies such as molecular electronics, solar energy harvesting, thermoelectrics, sensing, and other applications \cite{V15,Fe4,PhysRevLett.76.3830}. The Anderson model can be used to study impurities adsorbed on  surfaces \cite{PhysRevB.92.245135} and appears as an impurity model in the Dynamic Mean Field Theory\cite{RevModPhys.68.13}. \\
{\em Solution method:}\\
The OpenMP and MPI parallelized versions of the finite temperature Lanczos diagonalization method is used to diagonalize Hamiltonian matrix and to compute observables.
\end{small}

\section{Introduction}
Further progress in material science is connected with the development of appropriate theoretical concepts and methods to treat realistic modern materials and devices taking their atomic structure, chemical composition, electronic and magnetic properties fully into account.
Two of the basic quantum models for systems with strong electron-electron correlations are the Hubbard model\cite{Hubbard_1963}  and the Anderson impurity model\cite{PhysRev.124.41}, which can be used to simulate lattice problems or an impurity in metal respectively.

At the moment, there are a number of well-developed numerical techniques one can use to solve these quantum electron models. For instance, many interesting and promising results were obtained by using QMC-type methods such as continuous-time quantum Monte Carlo method \cite{PhysRevB.72.035122}.
Since the main computational task is a sampling of a complex integral, these methods are ideally suited for parallelization. However, there is a fundamental problem of the QMC solvers called the sign problem, which can occur for models with non-diagonal Coulomb interaction matrix, lattice problem away from half-filling or when the simulation temperature is rather low \cite{PhysRevB.80.155132}. 
 
Alternatively, truncating the infinite Hilbert space by solving a finite lattice problem or by discretizing an infinite bath with a finite set of energy levels allows one to use exact diagonalization techniques to treat the Anderson Hamiltonian. Such a method allows to diagonalize the electronic Hamiltonian for different geometries of lattice cluster or with different forms of the on-site Coulomb matrix \cite{Liebsch2011,PhysRevB.66.081104}. Another advantage of the exact diagonalization method  is that it provides a natural way to calculate real-frequency correlation functions such as one- and two-particle Green's functions at finite temperatures.

In this work, we present the parallel Exact diagonalization library for solving the eigenvalue
problem of the Hubbard model or Anderson impurity model on distributed-memory and shared memory computing systems.

\section{Exact diagonalization of finite quantum electron models}
The Hamiltonian of the many quantum electron model can be expressed as the sum of local (diagonal) term and non-diagonal hopping term as follows:
\begin{equation}
\mathcal{H} = \mathcal{H}_{loc} + \mathcal{H}_{hop}.
\end{equation}
For example in case of Hubbard model \cite{Hubbard_1963} $\mathcal{H}_{loc} = \sum\limits_i U_i n_{i \uparrow}n_{i \downarrow} - \sum\limits_{i \sigma} \mu_i n_{i \sigma}$ and $\mathcal{H}_{hop} = \sum\limits_{\langle i,j\rangle \sigma} t_{ij} c^\dagger_{i \sigma} c_{j \sigma},$
where
  $U_i$ is Coulomb potential on site $i$;
  $\mu_i$ -- chemical potential on site $i$;
  $t_{ij}$ -- hopping integral between sites $i$, $j$.
  $c^{(\dagger)}_{i \sigma}$ -- annihilation (creation) operator of electron with spin direction $\sigma$ on $i$-th site. $n_{i \sigma} = c^\dagger_{i \sigma} c_{i \sigma}$ -- occupation number, number of electrons on the site. 

The first step of exact diagonalization algorithm is to represent a Hamiltonian operator as a matrix. Despite the fact that for most quantum electron models this matrix is very sparse (99\% of matrix elements being zeroes) the dimension still grows exponentially $M = 2^{2 N_s}$ in occupation number space
$\ket{n_{1 \uparrow}, \dots, n_{N_s \uparrow} | n_{1 \downarrow}, \dots, n_{N_s \downarrow}}$, where $N_s$ is the number of electron levels in the studied quantum electron model. The exponential growth of basis size puts serious restriction on lattice size. 

\begin{figure}[htp]
\centering
\includegraphics[width=0.4\textwidth]{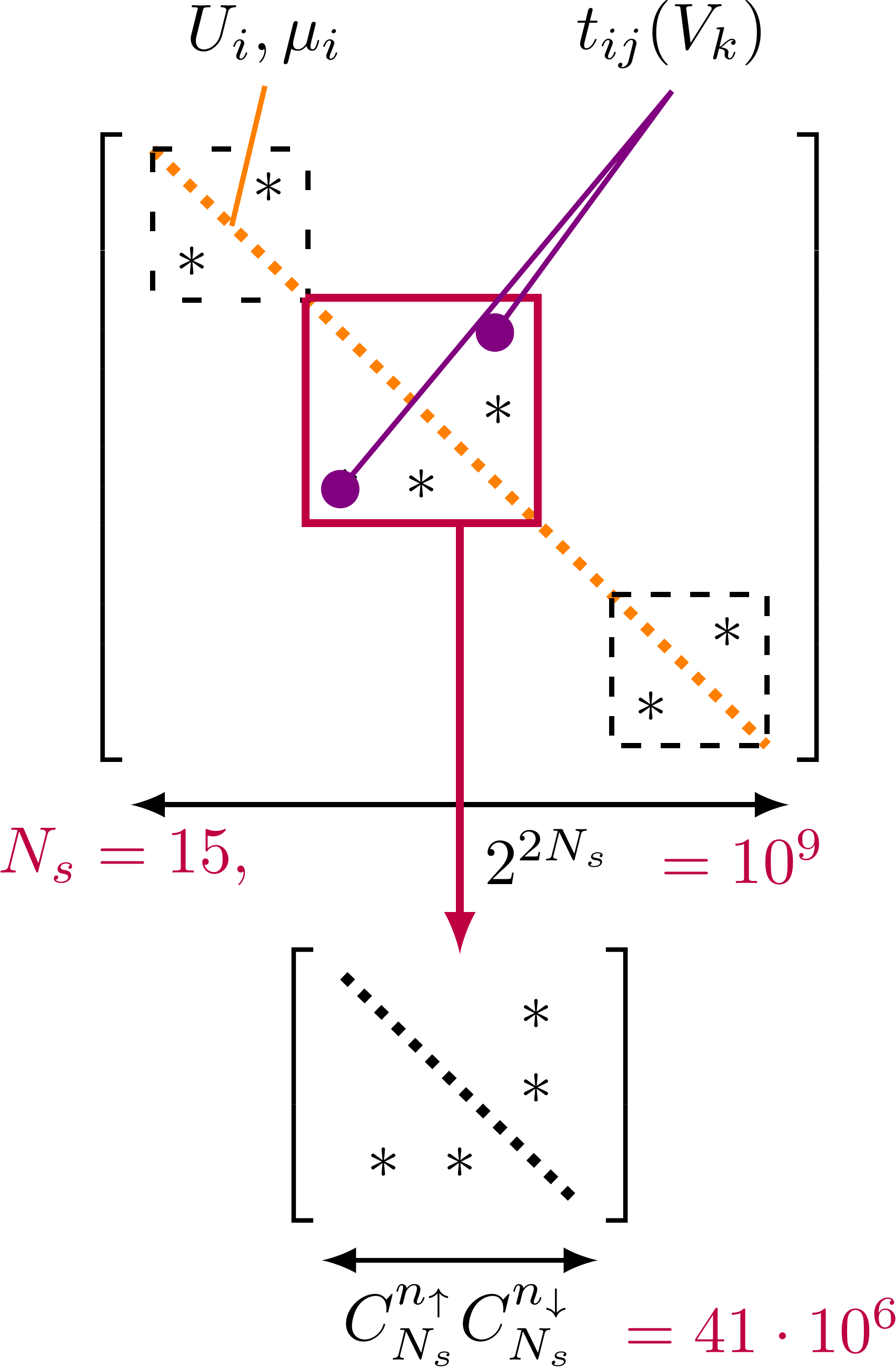}
\caption{Block structure of the $S^z$-symmetric model Hamiltonian matrix.}
\label{BlockMatrix}
\end{figure}

Based on the particle and spin conservation properties of the Hubbard model Hamitonian the matrix assumes block-diagonal form
and the blocks -- so called \emph{sectors} -- of much smaller dimension $M_{n_\uparrow n_\downarrow} = C_{N_s}^{n_\uparrow} \cdot C_{N_s}^{n_\downarrow}$,
where $C_n^k$ is the number of combinations of k from n elements.
The sectors can be diagonalized separately as matrices of local Hamiltonians for fixed total occupation numbers $n_\uparrow = \sum_{i=1}^{N_s} \hat{n}_{i \uparrow}$, $n_\downarrow = \sum_{i=1}^{N_s} \hat{n}_{i \downarrow}$ (see Fig.(\ref{BlockMatrix})).
The size difference for $N_s = 15$ in case of half-filling is 1.6 orders of magnitude, which is considerable, yet the amount of data and matrix sparsity (99\% of the elements are zeroes for both matrices) remain high enough to cause difficulties\cite{Lin}. In the case of the Lanczos diagonalization algorithm the main procedure is the Krylov subspace construction that requires a matrix-vector product operation.

\section{Exact diagonalization of single multi-orbital impurity Anderson Model}
The multi-orbital impurity Anderson model can be written in the following general form:
\begin{align}
H =  \sum_{p \sigma} \epsilon_{p} c^{+}_{p \sigma} c_{p \sigma} 
+& \sum_{i \sigma}   (\epsilon_{i} - \mu) n_{i \sigma} +  \sum_{i p \sigma} ( V_{i p} d^{+}_{i \sigma} c_{p \sigma} + H.c.) + \nonumber \\
 +& \frac{1}{2} \sum_{\substack {ijkl\\ \sigma \sigma'}}  U_{ijkl} d^{+}_{i \sigma} d^{+}_{j \sigma'} d_{l \sigma'} d_{k \sigma}.  
\end{align}
Here $\epsilon_{i}$ and $\epsilon_{p}$ are energies of the impurity and bath states, $d^{+}_{i \sigma}$ and $c^{+}_{p \sigma}$ are the creation operators for impurity and surface electrons, $V_{ip}$ is a hopping between impurity and surface states, $U_{ijkl}$ is the Coulomb matrix element and the impurity orbital index $i$ ($j$, $k$, $l$) runs over the $d-$ states. Depending on the problem we solve the bath can correspond to either an effective Weiss field (DMFT) or, for instance, metallic surface states (adatom on a substrate).

\section{Storage formats}
\subsection{Spin-resolved Hamiltonian storage format}
Since the hopping Hamiltonian does not contain hopping between different spins it can be decompose into two parts for each spins as follows:
\begin{equation}
\mathcal{H}_{hop} = \mathcal{H}_{\uparrow} \oplus \mathcal{H}_{\downarrow} = \mathcal{H}_{\uparrow} \otimes \mathcal{I}_{\downarrow} + \mathcal{I}_{\uparrow} \otimes \mathcal{H}_{\downarrow},
\label{decouple}
\end{equation}
where $\mathcal{I}_{\sigma}$ is the identity matrix with the same dimension as $\mathcal{H}_{\sigma}$, and can be stored separately. Since dimension of $\mathcal{H}_{\sigma}$ is much smaller than the original Hamiltonian matrix, the only problem is to store the eigen-vectors since the Hilbert space still grows exponentially. To deal with this issue in this library we implement the distributed storage of the vector as will be described in the next subsection.

\subsubsection{MPI parallelization}
In this library for solving the eigenvalue problem we use a parallel version of the implicitly restarted Arnoldi algorithm library \cite{arpack}, which requires implementation of the matrix-vector products. In the case of a matrix decoupled into diagonal and two off-diagonal matrices parts (See Eq. \ref{decouple}) this operation can be performed by three separate operations:
\begin{align}
H \left(\begin{matrix}x_1\\x_2\\\ldots\\x_3\end{matrix}\right) = H_{loc}\left(\begin{matrix}x_1\\x_2\\\ldots\\x_3\end{matrix}\right) +
\left(\begin{matrix}H_{\downarrow}x^{(\downarrow)}_1\\H_{\downarrow}x^{(\downarrow)}_2\\\ldots\\H_{\downarrow}x^{(\downarrow)}_3\end{matrix}\right) +
H_{\uparrow}\left(\begin{matrix}x^{(\downarrow)}_1\\x^{(\downarrow)}_2\\\ldots\\x^{(\downarrow)}_3 \end{matrix}\right),
\label{distribution}
\end{align}
where $x^{(\downarrow)}_{i}$ is an $i-$sub-vector of initial vector $x$ with a dimension of $dim(H_{\downarrow})$. It is clear to see vector can be simply distributed along different processors by integer numbers of $x^{(\downarrow)}_{i}$ sub-vectors. The only operation that needs to perform inter-processor communication is the last term in the right part of the Eq.\ref{distribution}. In this case we can overlap communications and computations by using one-sided MPI communications:
\begin{cpplisting}
// Perform initial synchronization
MPI_Win_fence(MPI_MODE_NOPRECEDE, _win);
// Initiate remote data transfer for up-spin term
MPI_Get(...)
// Compute diagonal contribution.
...
// Compute down-spin contribution
...
// Perform final synchronization
MPI_Win_fence(MPI_MODE_NOSUCCEED | MPI_MODE_NOPUT | MPI_MODE_NOSTORE, _win);

// Compute up-spin contribution and off-diagonal interactions contribution


\end{cpplisting}
\subsubsection{Scaling properties}
Table \ref{scaling} shows how the program scales with the number of MPI processes the two problems with different matrixs size on the University of Michigan high-performance cluster. We see that the computation time behaves like $1/0.8N_{p}$. The principal reason for this behavior is that a larger number of processes leads to more inter-node communication compared to mostly intra-node communication for a small number of processes. One  way to improve the scalability is to dedicate a single core for communications. Work in this direction is currently in progress.

\begin{table}[ht]
\centering
\begin{tabular}{p{3cm}|c | c }
\hline
\hline
Processes & 41409225 & 64128064 \\
\hline
36 & 0.91 sec & 2.47 sec \\
72 & 0.55 sec & 1.41 sec \\
\hline
\end{tabular}
\caption{Wall-clock time used per iteration as a function of the number of MPI processes on various matrix sizes.}
\label{scaling}
\end{table}

\subsection{Signs-only Compressed Row storage format}
In the new sparse matrix format -- \emph{<<Signs Only Compressed Row Storage>>} (SOCRS) -- we attempt to balance the time-efficiency of CRS with the size-efficiency of on-the-fly generation.

According to estimation, the off-diagonal elements of the Hamiltonian matrix stored in CRS format make the largest contribution to memory footprint.
The absolute values of inter-site hoppings $t_{ij}$ are readily available as an adjacency matrix, normally read once on solver initialization, whereas sign depends (for Fermi-Dirac statistics) on the number of sites occupied to the left of the changed state:
\begin{align*}
 \hat{c}_j \ket{m} &= (-1)^s\ket{\dots},  \\
  s &= \sum_{i \le j} n_i, \nonumber
\end{align*}
which means their on-the-fly calculation would considerably increase SpMV time. In this situation it is sensible to store only signs in sparse format, one sign per bit. The dense diagonal can be stored in a separate vector. Scanning the adjacency matrix makes the number of elements in the row known, thus making \texttt{row\_ptr} of CRS redundant.

\begin{figure}[thp]
 \centering
  \includegraphics[width=0.5\textwidth]{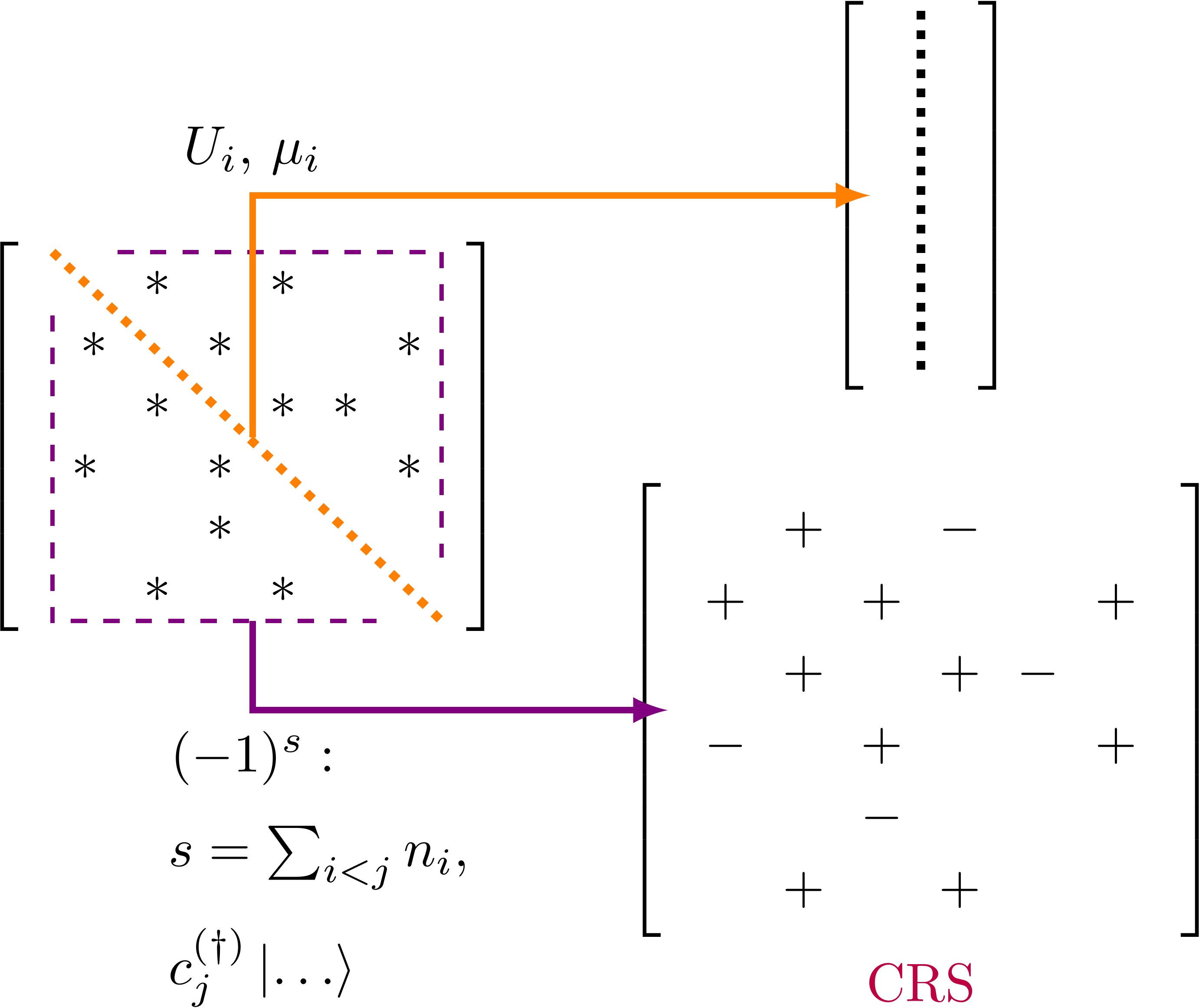}
  \caption{
   SOCRS -- suggested variation of CRS format.
   Signs of off-diagonal elements are stored in compressed format, the diagonal is stored separately.
  }
  \label{SOCRS}
\end{figure}

\section{Program description}
The EDLib library is designed to solve the exact diagonalization problem for electronic quantum Hamiltonians. For a large matrix cases we use parallelization by means of MPI or OpenMP (depending storage type).
The program is written in standard C++11 and distributed as a template library. The program has been checked using GNU, Intel and Clang C++ compilers. The test run has also been checked on the University of Michigan high-performance high-performance computing (HPC) cluster.

\begin{figure}[thp]
 \centering
  \includegraphics[width=0.9\textwidth]{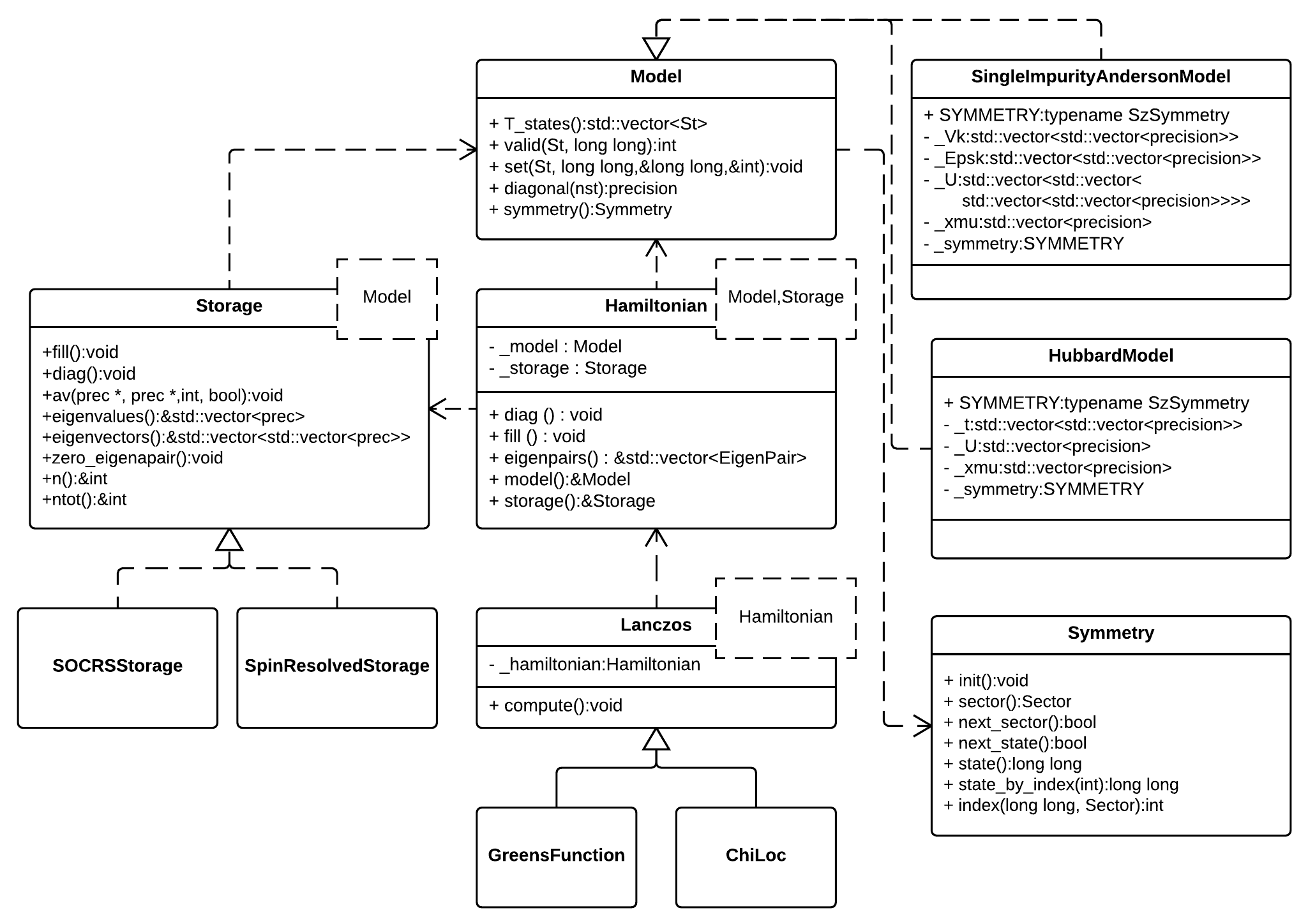}
  \caption{
   UML diagram of classes implemented in EDLib library.
  }
  \label{UML}
\end{figure}
\subsection{Class diagram}
The structure of the EDLib library can be represented by the UML diagram shown at Fig. (\ref{UML}). The main class of the presented library is the $Hamiltonian$ which is parametrized by type of $Model$ and type of $Storage$. The main method for exact diagonalization is $diag$ that mostly delegate the work to the specified $Storage$ class. The Green's functions ($GreenFunction$ class) are computed by the Lanczos continued fraction method ($Lanczos$ class).

\subsection{Description of the input data}
Input data to the EDLib library take the form of (1) parameter file, (2) Model specific HDF5 file. Table \ref{params} represents the complete list of the parameters that can be defined in the parameter file. The structure of HDF5 files is model specific and for its generation we provide Python scripts for each represented model.
\begin{table}[ht]
\centering
\begin{tabular}{p{6cm}|c}
\hline
\hline
Parameter name & Description\\
\hline
NSITES &Number of sites\\
NSPINS & Number of spins\\
INPUT\_FILE& HDF5 input file\\
\hline
\multicolumn{2}{c}{Storage parameters}\\
\hline
storage.MAX\_SIZE& Number of eigenvalues to find\\
storage.MAX\_DIM & Number of eigenvalues to find\\
storage.EIGENVALUES\_ONLY& Compute only eigenvalues\\
spinstorage.ORBITAL\_NUMBER& Number of orbitals with interaction\\
\hline
\multicolumn{2}{c}{ARPACK parameters}\\
\hline
arpack.SECTOR& Read symmetry sectors from file\\
arpack.NEV& Number of eigenvalues to find\\
arpack.NCV& Number of convergent values \\
\hline
\multicolumn{2}{c}{Lanczos parameters}\\
\hline
lanc.NOMEGA& Number of Matsubara frequencies \\
lanc.NLANC& 100, "Number of Lanczos iterations\\
lanc.BETA& 10.0, "Inverse temperature \\
lanc.BOLTZMANN\_CUTOFF& Cutoff for Boltsman factor \\
\hline
\multicolumn{2}{c}{single impurity Anderson Model}\\
\hline
siam.NORBITALS& Number of impurity orbitals\\
\hline
\end{tabular}
\caption{Input parameters description}
\label{params}
\end{table}

\section{Prerequisites and Installation}
To build the EDLib library, any recent C++11 compiler can be used; the libraries have been tested with GCC \cite{gcc} 4.6 and above, Intel C++ 15 and above and Clang \cite{clang} 3.4 and above.

The library depends on the following packages:

\begin{itemize}
 \item The \verb|CMake| build system \cite{cmake} of version 2.8.12 and above.
 \item The \verb|ALPSCore| libraries \cite{Gaenko2016} of version 0.54.0 and above.
 \item The \verb|ARPACK| library \cite{arpack} of opencollab arpack-ng \cite{arpack-ng} version 3.5.0 and above. 
\end{itemize}

To use optional distributed or shared memory parallel capabilities (the support depends on chosen matrix storage format), an MPI implementation supporting standard 2.1 \cite{mpi-2.1} and above, or OpenMP version 3.0 \cite{openmp08} must be enabled by \texttt{USE\_MPI} or \texttt{USE\_OPENMP}, respectively.

The installation of the EDLib library follows the standard procedure for any CMake-based package.
The first step is to download the EDLib source code. Assuming that all above mentioned prerequisite software is installed, the installation consists of running CMake from a temporary build directory, as outlined in the shell session example with MPI support below:
\begin{bashlisting}[emph={tar,mkdir,cmake,make},emphstyle={\color{darkgreen}}]
git clone https://github.com/Q-Solvers/EDLib.git
mkdir EDLib-build&& cd EDLib-build

cmake \
 -DCMAKE_INSTALL_PREFIX=${HOME}/local/EDLib/ \
 -DALPSCore_DIR=${HOME}/local/ALPSCore/share/ALPSCore/ \
 -DARPACK_DIR=${HOME}/local/arpack-ng/lib/ \
 -DUSE_MPI=TRUE\
 ../EDLib

make
make test
\end{bashlisting}
The command at line {1} will download the latest source code from github; at line 5 the destination install directory of the EDLib libraries is set (\verb|${HOME}/local/EDLib| in this example).

\subsection{Citation policy and contributing}
EDLib, as an application of ALPSCore, is an open source project and we encourage feedback and contributions from the user community. Issues should be reported exclusively via the GitHub website at \url{https://github.com/Q-solvers/EDLib/issues}. For contributions, we recommend to use the pull request system on the GitHub website. Before any major contribution, we recommend coordination with the main EDLib developers.
We kindly request that the present paper be cited in any published work using the EDLib library as well as the ALPSCore library on which the presented library is based also be cited\cite{Gaenko2016}. This helps the EDLib and ALPSCore developers to better keep track of projects using the library and provides them guidance for future developments. 

\section{Examples} 
\begin{figure}[htp]
\centering
\includegraphics[width=0.4\textwidth]{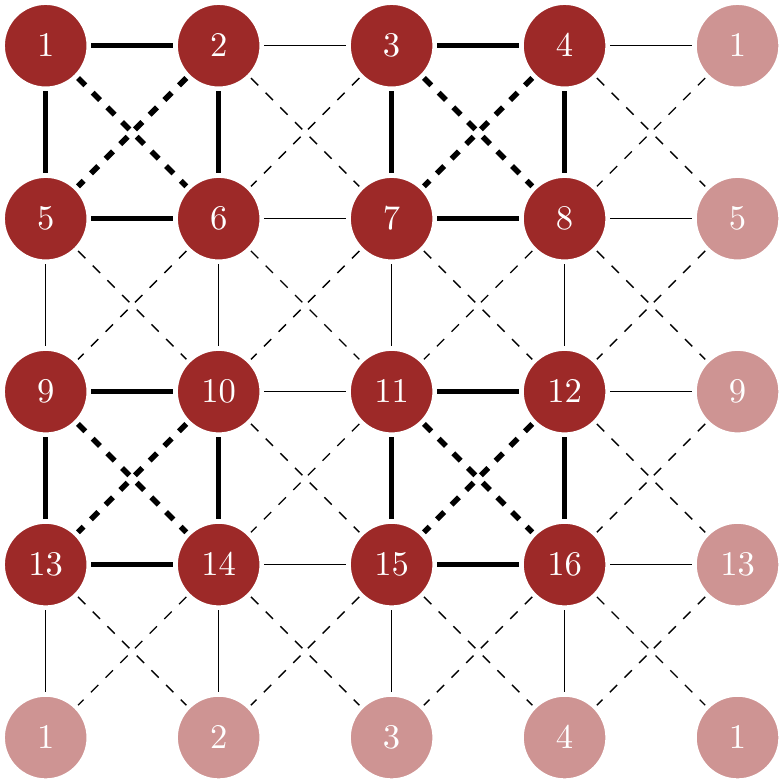}
\caption{Schematic representation of Hubbard 4-site plaquettes.}
\label{HubbardCluster}
\end{figure}

To show the ability of the presented library we consider two problems. The first problem is the groundstate calculation of isolated 4x4 Hubbard cluster. And the second one is the ground state electronic configuration of the single $Co$ impurity adsorbed on the $Pt(111)$ surface.

\subsection{Finite Hubbard cluster diagonalization}

The Hubbard 4-site plaquette is a minimal and generic electronic-structure model of cuprate superconductors suggested in \cite{PhysRevB.94.125133} --  which demonstrates critical behavior for certain doping.
Its properties have been studied in isolation, in the bath and in the Bethe lattice.
We study an isolated system of four such plaquettes with periodic boundary conditions. The schematic representation of the cluster is presented in the Fig. (\ref{HubbardCluster}). The solid and dashed lines correspond to nearest neighbour hopping $t$ and second nearest neighbour hopping $t'$ respectively. For the present calculation the following parameters have been chosen: $t=1.0 eV$, $t'=-0.3 eV$, $U=6.0 eV$ and $\mu=0.54 eV$. The resulting lowest energy is $E_0=-22.6421 eV$.

\begin{figure}[htp]
\centering
\includegraphics[width=0.45\textwidth]{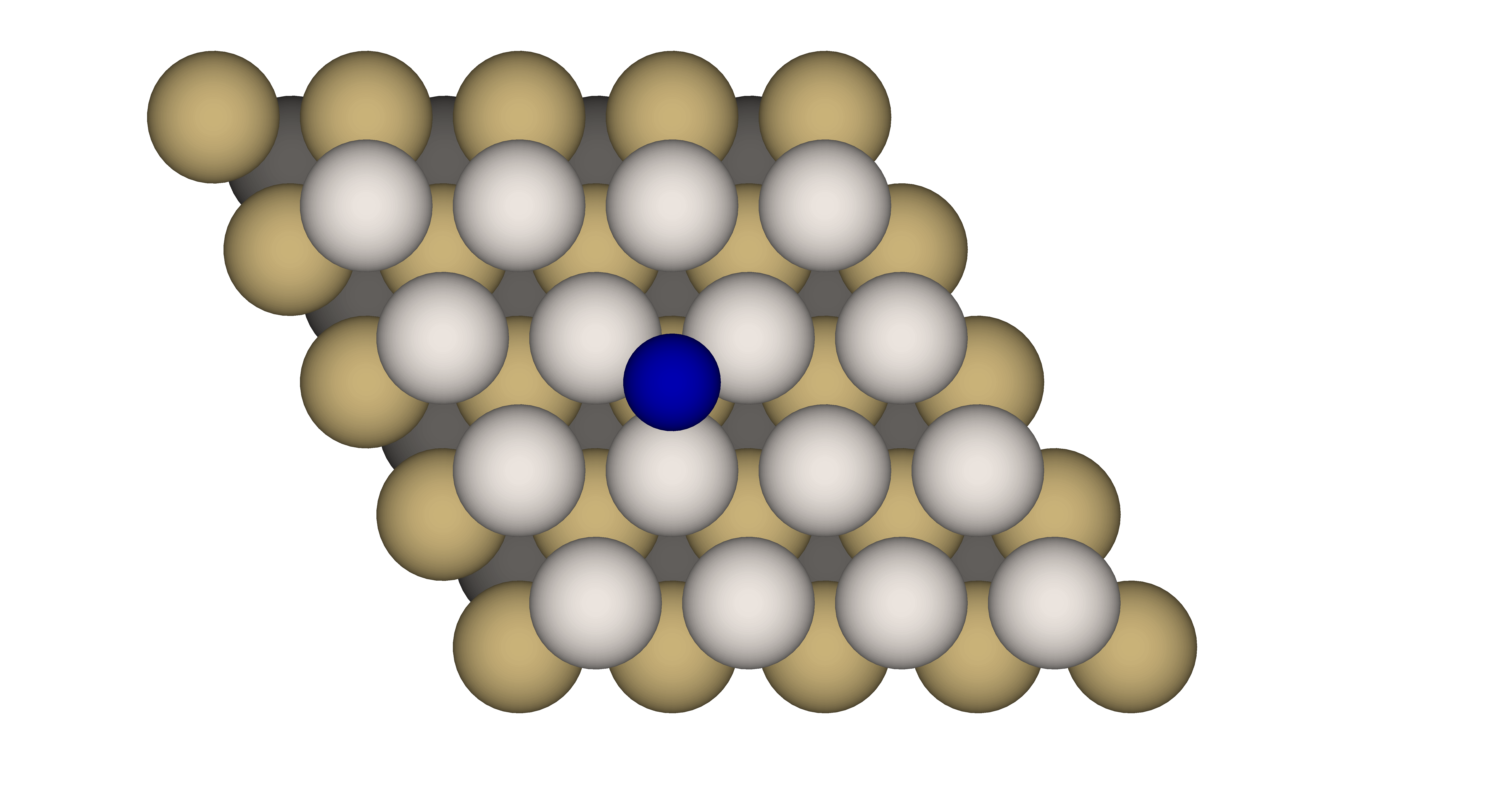}
\includegraphics[width=0.45\textwidth]{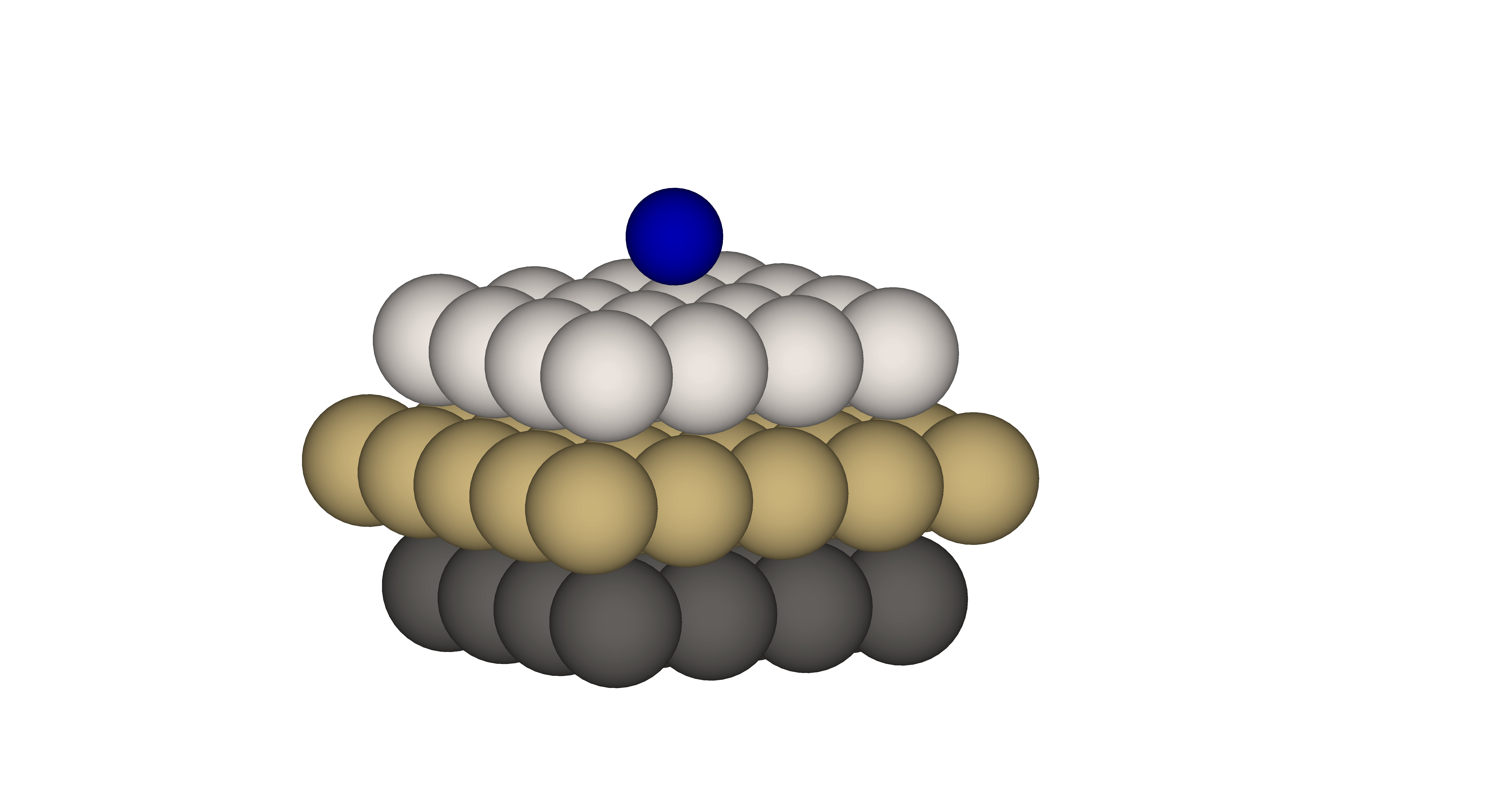}
\caption{The schematic representation of $hcp$ position of Co adatom adsorbed on the Pt(111) surface. The blue sphere shows cobalt atom and gray spheres correspond to Pt.}
\label{CoPt}
\end{figure}
\subsection{Co adatom on the Pt(111) surface}
The electronic and transport properties of the single transition metal adsorbed on the various type of surface play a crucial role in the proper description of giant magnetic anysotropy
\cite{Gambardella16052003} or Kondo physics \cite{Kondo,1367-2630-12-6-063040,FMeier}. In this simulation we present results for electronic configuration of the ground state of the single cobalt adatom adsorbed on the Pt(111) surface by means of the Anderson impurity model. The model parameters are obtained from first principles calculation. The experimental value of lattice constant for the bulk fcc Pt is 3.92 {\AA} \cite{PhysRevB.77.184425,PhysRevB.82.193403}. Since the relaxation for different stackings, $fcc$ and $hcp$, as shown in previous study \cite{PhysRevLett.102.257203} does not show much difference, we perform simulation for $hcp$ position of cobalt adatom as presented in Fig. \ref{CoPt}. 

\begin{table}[ht]
\centering
\begin{tabular}{p{3cm}|c | c }
\hline
\hline
Orbital & $V_k$, eV & $\epsilon_k$, eV \\
\hline
$xy$, $x^2-y^2$ & 0.56434; 0.68392; 0.29519  & -2.37325; -0.87328; 2.01265  \\
$xz$, $yz$ & 0.81892; 0.99136 & -3.15496; -1.69066  \\
$3z^2 - r^2$ & 0.77347; 0.79785  &  -5.59842; -2.95325  \\
\hline
\end{tabular}
\caption{Descitized bath parameters for orbitals of different symmetries.}
\label{Bath}
\end{table}
For the Anderson Impurity model we choose 5 orbitals for $d-$states of cobalt adatom, two orbitals in the bath for each $xz$,$yz$ and $3z^2-r^2$ cobalt orbitals and three orbitals in the bath for each $xy$ and $x^2-y^2$ cobalt orbitals. Based on the spin symmetries and block-diagonal structure of the Hamiltonian matrix the dimension of the largest block is about $\approx 590\times 10^6$. For the present calculations we choose the following parameters: $U=6.6$ eV, $J_H=0.9$ eV, $\mu=44.44$ eV, and the bath parameters are presented in the Table \ref{Bath}. 
The interaction part of the Hamiltonian is expressed by using Slater integral representation of the full rotational invariant Coloumb interaction tensor \cite{Slater} with $F^0 = U$, $F^2 = 14J_H/(1+0.625)$ and $F^4 = 0.625F^2$ \cite{PhysRevB.42.5459}. We perform diagonalization of each symmetry sector to find the electronic configuration of the ground state. 

The simulation is performed on Edison Cray machine and takes about 1600 core-hours on 10 nodes with maximum memory requirement about 10 Gb per node. The resulting electronic configuration is presented in Table \ref{ResCo}, in addition we present the lowest energy for the half-filled states.
\begin{table}[ht]
\centering
\begin{tabular}{p{1.5cm}|c | c |c |c }
\hline
\hline
$\Delta E$, eV & n$_{\uparrow}$ & n$_{\downarrow}$ &  Sector size & Major contribution to g.s. \\
\hline
0.0	&12	&15 & 841568 & \textbf{\sout{$\downarrow$} \sout{$\uparrow\downarrow$} \sout{$\downarrow$} \sout{$\uparrow\downarrow$} \sout{$\downarrow$}} \\
0.0	&13	&14 & 1618400 & \textbf{\sout{$\uparrow$} \sout{$\uparrow\downarrow$} \sout{$\downarrow$} \sout{$\uparrow\downarrow$} \sout{$\downarrow$}}
{\footnotesize{+}} \textbf{\sout{$\downarrow$} \sout{$\uparrow\downarrow$} \sout{$\downarrow$} \sout{$\uparrow\downarrow$} \sout{$\uparrow$}} {\footnotesize{+}}
\textbf{\sout{$\downarrow$} \sout{$\uparrow\downarrow$} \sout{$\uparrow$} \sout{$\uparrow\downarrow$} \sout{$\downarrow$}} \\
0.0	&14	&13 & 1618400 & \textbf{\sout{$\downarrow$} \sout{$\uparrow\downarrow$} \sout{$\uparrow$} \sout{$\uparrow\downarrow$} \sout{$\uparrow$}}
{\footnotesize{+}} \textbf{\sout{$\uparrow$} \sout{$\uparrow\downarrow$} \sout{$\uparrow$} \sout{$\uparrow\downarrow$} \sout{$\downarrow$}} {\footnotesize{+}}
\textbf{\sout{$\uparrow$} \sout{$\uparrow\downarrow$} \sout{$\downarrow$} \sout{$\uparrow\downarrow$} \sout{$\uparrow$}} \\
0.0	&15	&12 & 841568 & \textbf{\sout{$\uparrow$} \sout{$\uparrow\downarrow$} \sout{$\uparrow$} \sout{$\uparrow\downarrow$} \sout{$\uparrow$}} \\
\multicolumn{5}{@{}c}{\dotfill} \\
13.471& 9 & 8 & 590976100 & \\
13.471& 8 & 9 & 590976100 & \\
\hline
\end{tabular}
\caption{The resulting electronic configuration for exact diagonalization study of Co adatom adsorbed on Pt(111).}
\label{ResCo}
\end{table}

\section{Summary}
We have presented the free software EDLib, an implementation of the Exact diagonalization solver for Anderson Impurity and finite Hubbard models.
Further developments (e.g., support for complex Hamiltonians, other measures) are planned for a future release.

\section*{Acknowledgment}
We thank E. Gull and A. I. Lichtenstein for useful discussions. 
SI was supported by the Simons collaboration on the many-electron problem and by Act 211 Government of the Russian Federation, contract No. 02.A03.21.0006, M.D. acknowledge support from Deutsche Forschungsgemeinschaft via Project SFB 925.
This research used resources of the National Energy Research Scientific Computing Center, a DOE Office of Science User Facility supported by the Office of Science of the U.S. Department of Energy under Contract No. DE-AC02-05CH11231.

\bibliographystyle{elsarticle-num}
\bibliography{edlib}

\end{document}